%% file: Nbody.tex
\documentclass[runningheads]{llncs}
\usepackage[utf8]{inputenc}
\usepackage{graphicx}
\usepackage{hyperref}
\usepackage{amssymb}
\usepackage{booktabs}
\usepackage{amsmath}
\usepackage{subcaption}
\usepackage{picins,caption}
\usepackage{mathtools}
\usepackage[linesnumbered,ruled,vlined]{algorithm2e}
\usepackage{listings}
\usepackage{booktabs}
\usepackage{multirow}
\usepackage[table,xcdraw]{xcolor}
\usepackage{wrapfig}

\def\accd{\mathsf{nsb}}
\def\acce{\mathsf{nsb_e}}

\def\ufp{\mathsf{ufp}}
\def\ulp{\mathsf{ulp}}
\def\ufpe{\mathsf{ufp_e}}
\def\ulpe{\mathsf{ulp_e}}

 \title{A Study of the Floating-Point Tuning Behaviour on the N-body Problem}
\author{Dorra Ben Khalifa\inst{1}\and  Matthieu Martel\inst{1,2}}
\authorrunning{Ben Khalifa et al.}
\institute{University of Perpignan, LAMPS laboratory, 52 Av. P. Alduy, Perpignan, France \and Numalis, Cap Omega, Rond-point Benjamin Franklin, Montpellier, France
 \email{\{dorra.ben-khalifa, matthieu.martel\}@univ-perp.fr}}
\begin{document}
\maketitle
\begin{abstract}
In this article, we apply a new methodology for precision tuning to the N-body problem. Our technique, implemented in a tool named POP, makes it possible to optimize the numerical data types of a program performing floating-point computations by taking into account the requested accuracy on the results. POP reduces the problem of finding the minimal number of bits needed for each variable of the program to an Integer Linear Problem (ILP) which can be optimally solved in one shot by a classical linear programming solver. The POP tool has been successfully tested on programs implementing several numerical algorithms coming from mathematical libraries and other applicative domains such as IoT. In this work, we demonstrate the efficiency of POP to tune the classical gravitational N-body problem by considering five bodies that interact under gravitational force from one another, subject to Newton’s laws of motion. Results on the effect of POP in term of mixed-precision tuning of the N-body example are discussed.
\smallskip
\end{abstract}
\input{introduction}
\input{related}
\input{running}
\input{pop}
\input{experim}
\input{conclu}
\bibliographystyle{splnc}
\bibliography{NbodyBib}
\end{document}

%% file: introduction.tex
 \section{Introduction}\label{sec1}

Reducing the precision of floating-point data, also called precision tuning, provides for many practitioners in High-Performance Computing (HPC) and related fields an opportunity for exploring trade-offs in accuracy and performance~\cite{LVMS19}. Compared to higher-precision data formats (e.g. \texttt{binary64} and \texttt{binary128}), lower-precision data formats result in higher performance for computational intensive applications such as lower resource cost, reduced memory bandwidth requirements and energy consumption: the performance of \texttt{binary32} operations on modern architectures is often at least twice as fast as the performance of \texttt{binary64} operations \cite{IEEE754}.

A number of tools has been developed to assist developers in exploring the trade-off between floating-point accuracy and performance~\cite{CBBSGR17,DHS18,GR18,KSWLB19,LHSL13,RGNNDKSBIH13}. A common purpose of these techniques is that they follow a try and fail strategy to reduce the precision with respect to an accuracy constraint: they create a program search space then tune either program variables or assembly instructions while the optimized data formats are dependent to the tuning inputs. If the accuracy of the results are not satisfying, some variables or instructions are removed from the search space and this process is applied repeatedly until a (locally) optimal solution is returned. In the best of cases of recent tools, the program is no longer treated as a black-box but an analysis of the source code and the runtime behaviour is performed in order to provide a customized search space classification and then to identify dependencies among floating-point variables.

Unlike existing approaches, POP~\cite{KM19,KM20,KM21,KMA19} implements a static technique based on a semantical modelling of the propagation of the numerical errors throughout the code. This results in generating a system of constraints whose minimal solution gives the best tuning of the program, furthermore, in polynomial time. The key feature of our method is to find directly the minimal number of bits needed, known as bit-level precision tuning, at each control point to get a certain accuracy on the results. Hence, it is not dependant of a certain number of data types (e.g. the IEEE754 formats~\cite{IEEE754}) and its complexity does not increase as the number of data types increases. In practical terms, by reasoning on the number of significant bits of the program variables and knowing the weight of their most significant bit thanks to a range analysis performed before the tuning phase (details in Section~\ref{sec4}), POP is able to reduce the problem to an Integer Linear Problem (ILP) which can be optimally solved in one shot by a classical linear programming solver (no iteration) $-$ we use GLPK~\cite{GLPK} in practice. The method scales up to the solver limitations and the solutions are naturally found at the bit level. Furthermore, POP implements an optimization to the previous ILP method. The purpose of this new method is to handle carry bits by being less pessimistic on their propagation throughout arithmetic expressions. By doing so, a second finer set of constraints is generated by POP and the problem does not remain any longer to a pure ILP problem ($\min$ and $\max$ operators are needed). Then we use policy iteration (PI) technique~\cite{CGGMP05} to find optimal solutions.
\paragraph{\textbf{Proposed Contributions}} In this article, we validate the efficiency of our approach on one of the oldest problem of modern physics, the N-body problem~\cite{Nbody}. An N-body simulation numerically approximates the evolution of a system of bodies that interact with one another through some type of physical forces, where $N$ presents the number of bodies in the system ($N=5$). The program implements a second order differential equation which needs to be solved to get a location of the bodies for a given timevalue. By varying the required accuracy by the user,  we show experimentally that POP succeeds in tuning the N-body program (original program $\simeq$ $330$ LOCs). As a result, the transformed program is guaranteed to use variables of lower precision with a minimal number of bits than the original one. Prior work on the precision of N-body simulations have been carried out for a long time~\cite{MakinoKF03}. Compared to other experiments carried out with POP~\cite{KM19,KM20}, the N-body example presents new difficulties, mainly more complex computations and a wide range of values with different magnitudes. 

 The different experimental evaluations presented in this article are the following. First, we measure the distance between the exact position of the bodies, \texttt{Jupiter}, \texttt{Saturn}, \texttt{Uranus} and \texttt{Neptune} (the $\texttt{Sun}$ position is fixed), computed with $500$ bits and the position computed with $n$ bits where $n= 11, 18, 24, 34 ,43$ and $53$ bits. These distances are given for each body with different time of simulation ($10$ and $30$ years). Second, we demonstrate on this example the ability of POP to generate an MPFR code~\cite{mpfr} with the new data types returned by the solver. Furthermore, we measure the global analysis time taken by POP and the execution time of the MPFR generated code and we prove that POP returns solutions in a few of a seconds. The global analysis time includes the time of the program evaluation, the range analysis determination, the constraint generation and their resolution by the solver.
\paragraph{\textbf{Roadmap}} The rest of this article is structured as follows. Section~\ref{sec2} describes the existing approaches for precision tuning. In section~\ref{sec3}, we discuss the technique behind POP on the illustrative example of N-Body problem. In section~\ref{sec4}, we point out the key contributions made by POP to tune floating-point programs. Section~\ref{sec5} presents a comprehensive evaluation and the experimental results of POP on the N-body problem and concluding remarks are discussed in Section~\ref{sec6}. 

%% file: related.tex
 \section{Related Work}\label{sec2} As we have discussed, the general areas for floating-point precision tuning have been receiving a lot of attention. In order to attain optimally lowered precision, such approaches are divided into two categories: formulating precision tuning as an optimization problem using a static performance and accuracy model, and  dynamically searching through different precisions to find a local optimum.
\paragraph{\textbf{Static Performance and Accuracy Model}} Prior work in static error analysis provides a foundation for rigorously determining what precisions are required to meet error constraints for particular closed form equations. In this context, the \textsc{FP-Taylor} tool~\cite{CBBSGR17} has proposed a method called Symbolic Taylor Expansions in order to estimate round-off errors of floating-point computations. Unlike dynamic tools, the precision allocation guarantees to meet the error target across all program inputs in an interval. Even so, \textsc{FP-Taylor} is not designed to be a tool for complete analysis of floating-point programs: conditionals and loops can not be handled directly. More recently, they have extended their work by performing a broad comparison of many error bounding analyses to ensure the mixed-precision tuning technique~\cite{fptuner}. Darulova et al.~\cite{DHS18} proposed a technique to rewrite programs by adjusting the evaluation order of arithmetic expressions prior to tuning. However, the technique is limited to rather small programs that can be verified statically. Basically, all the above methods suffer from scalability limitations and do not leverage community structure to guide the search. On the other hand, concerning scalability, POP generates a linear number of constraints and variables in the size of the analyzed program.
 \paragraph{\textbf{Dynamic Searching Applications}}
 \textsc{Precimonious}~\cite{RGNNDKSBIH13} is a dynamic automated search based tool that leverages the LLVM framework to tweak variable declarations to build and prototype mixed-precision configurations. It aims at finding the \textit{1-minimal} configuration, i.e., a configuration where
 changing even a single variable from higher to lower precision would cause the configuration to cease to be valid. A valid configuration is defined as one in which the relative error in program output is within a given threshold and there is a performance improvement compared to the baseline version of the program. However, it does not use any knowledge on the structure of the program to identify potential variables of interest.     Lately, a new solution called \textsc{HiFPtuner}~\cite{GR18}, which is an extension Of \textsc{Precimnious}, uses dependence analysis and edge profiling to enable a more efficient hierarchical search algorithm for mixed-precision configurations. As with other dynamic tuners,  \textsc{HiFPtuner}’s configurations are dependent on the tuning inputs, and no accuracy guarantee is provided for untested inputs. \textsc{Craft}~\cite{LHSL13} is a framework that uses binary instrumentation and modification to build mixed-precision configurations of existing binaries that were originally developed to use only double-precision. \textsc{Stoke}~\cite{stoke} is a general stochastic optimization and program synthesis tool to handle floating-point computation. Their algorithm applies a variety of program transformations, trading bit-wise precision for performance to enhance compiler optimization on floating-point binaries. Another dynamic tool is \textsc{Adapt}~\cite{adapt}. It uses the reverse mode of algorithmic differentiation to determine how much precision is needed in a program inputs and intermediate results in order to achieve a desired accuracy in its output, converting this information into precision recommendations. Following the idea of program transformation, the \textsc{Ampt-Ga} tool~\cite{KSWLB19} selects application-level data precisions to maximize performance while satisfying accuracy constraints. \textsc{Ampt-Ga} combines static analysis for casting-aware performance modeling
      with dynamic analysis for modeling and enforcing precision constraints.
POP only focuses on the precision tuning problem. Hence, the input codes are taken as-is and we do not modify them. However, POP is compatible with other tools for program transformation for numerical accuracy ~\cite{DamoucheM18,SFNPT21}. Typically, these tools re-order the computations to make them more accurate in the computer arithmetic. For example, for sums, numbers will be added in increasing order of magnitude.

%% file: running.tex
 \lstset{language=Java, numbers=left, showspaces=false,
   showstringspaces=false, tabsize=2, breaklines=true,
   xleftmargin=5.0ex,
    numberstyle=\scriptsize,numbersep=0pt
}

\section{Running Example}\label{sec3}


\pichskip{15pt}
\parpic[r][b]{%
  \begin{minipage}{6.5cm}
    \includegraphics[width=6.5cm]{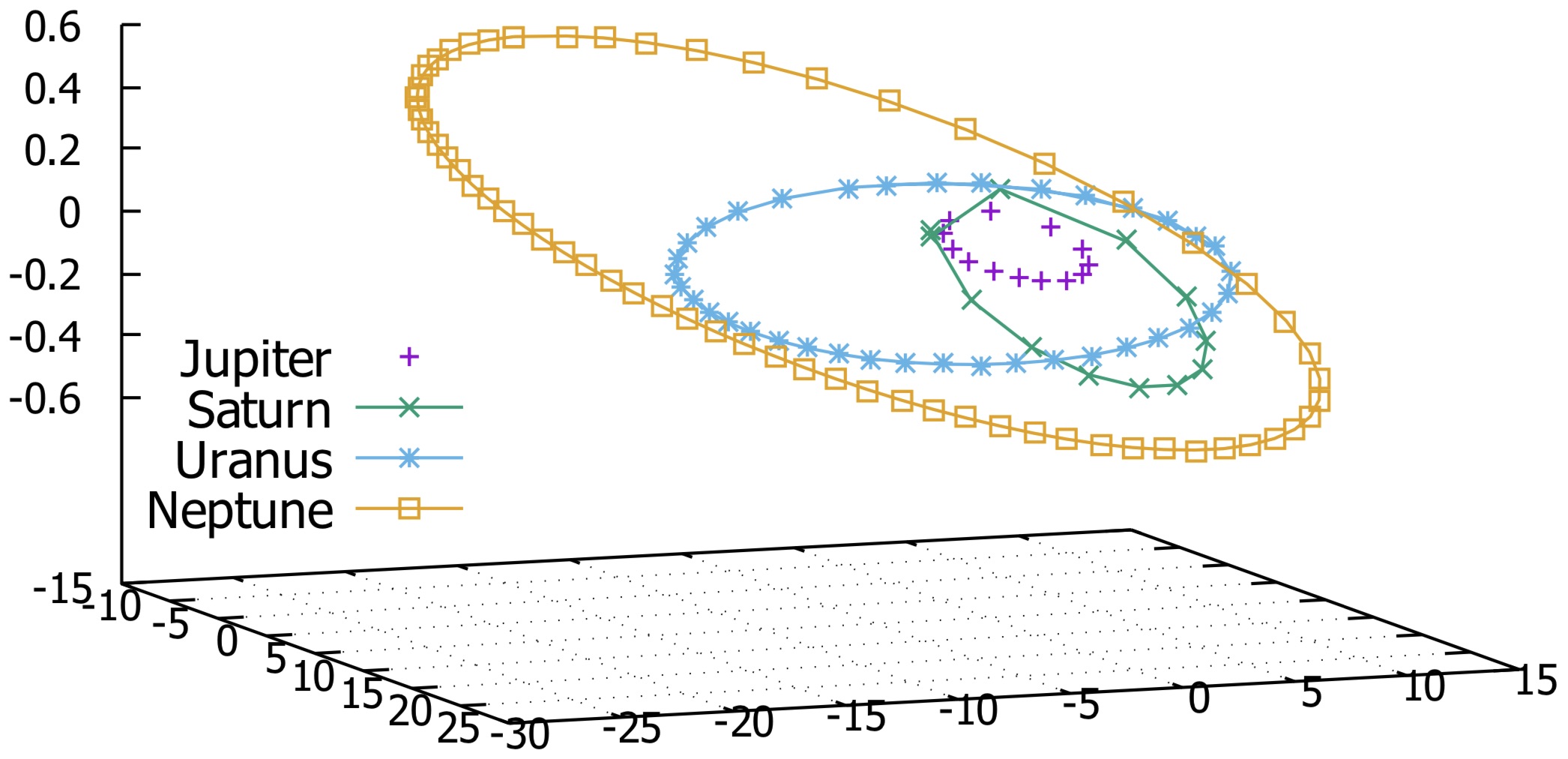}
\captionof{figure}{\label{orbites} Simulated movement of the bodies.}
 \end{minipage}
}
\begin{figure}[!h]
\hrule
\vspace{0.2cm}
\scriptsize
\centering
\begin{tabular}{lr}\tt
\begin{lstlisting}[mathescape]
 days_per_year$^{\ell_{11}}$ = 365.24$^{\ell_{10}}$;
 dt$^{\ell_{13}}$ = 0.01$^{\ell_{12}}$;
 t$^{\ell_{15}}$ = 0.0$^{\ell_{14}}$;
 t_max$^{\ell_{17}}$ = 1000.0$^{\ell_{16}}$;
              [...]
 xJupiter$^{\ell_{39}}$ = 4.8414316$^{\ell_{38}}$;
 vxJupiter$^{\ell_{48}}$ = 0.0016600767$^{\ell_{44}}$
  *$^{\ell_{47}}$ days_per_year$^{\ell_{46}}$;
 massJupiter$^{\ell_{63}}$ = 9.5479196E-4$^{\ell_{59}}$
 *$^{\ell_{62}}$ solar_mass$^{\ell_{61}}$;
 xSaturn$^{\ell_{65}}$ = 8.343367$^{\ell_{64}}$;
              [...]
 vxSaturn$^{\ell_{74}}$ = -0.002767425$^{\ell_{70}}$
 *$^{\ell_{73}}$ days_per_year$^{\ell_{72}}$;
 massSaturn$^{\ell_{89}}$ = 2.8588597E-4$^{\ell_{85}}$
 *$^{\ell_{88}}$ solar_mass$^{\ell_{87}}$;
              [...]
 while (t$^{\ell_{143}}$ <$^{\ell_{146}}$ t_max$^{\ell_{145}}$) {
   dx$^{\ell_{757}}$ = xJupiter$^{\ell_{753}}$ -$^{\ell_{756}}$ xSaturn$^{\ell_{755}}$;
   dy$^{\ell_{763}}$ = yJupiter$^{\ell_{759}}$ -$^{\ell_{762}}$ ySaturn$^{\ell_{761}}$;
   dz$^{\ell_{769}}$ = zJupiter$^{\ell_{765}}$ -$^{\ell_{768}}$ zSaturn$^{\ell_{767}}$;
   distance$^{\ell_{788}}$ = sqrt(dx$^{\ell_{771}}$ *$^{\ell_{774}}$ dx$^{\ell_{773}}$
   +$^{\ell_{780}}$ dy$^{\ell_{776}}$ *$^{\ell_{779}}$ dy$^{\ell_{778}}$ +$^{\ell_{786}}$ dz$^{\ell_{782}}$
   *$^{\ell_{785}}$ dz$^{\ell_{784}}$)$^{\ell_{787}}$;
   mag$^{\ell_{800}}$ = dt$^{\ell_{790}}$ /$^{\ell_{799}}$ distance$^{\ell_{792}}$ *$^{\ell_{795}}$
   distance$^{\ell_{794}}$ *$^{\ell_{798}}$ distance$^{\ell_{797}}$;
   vxJupiter$^{\ell_{812}}$ = vxJupiter$^{\ell_{802}}$ -$^{\ell_{811}}$
   dx$^{\ell_{804}}$ *$^{\ell_{807}}$ massSaturn$^{\ell_{806}}$ *$^{\ell_{810}}$ mag$^{\ell_{809}}$;
              [...]
   vxSaturn$^{\ell_{848}}$ = vxSaturn$^{\ell_{838}}$ +$^{\ell_{847}}$ dx$^{\ell_{840}}$
   *$^{\ell_{843}}$ massJupiter$^{\ell_{842}}$ *$^{\ell_{846}}$ mag$^{\ell_{845}}$;
              [...]
   xJupiter$^{\ell_{2602}}$ = xJupiter$^{\ell_{2595}}$ +$^{\ell_{2601}}$
    dt$^{\ell_{2597}}$ *$^{\ell_{2600}}$ vxJupiter$^{\ell_{2599}}$;
   xSaturn$^{\ell_{2683}}$ = xSaturn$^{\ell_{2676}}$ +$^{\ell_{2682}}$ dt$^{\ell_{2678}}$
    *$^{\ell_{2681}}$ vxSaturn$^{\ell_{2680}}$;
              [...]
   t$^{\ell_{2707}}$ = t$^{\ell_{2703}}$ +$^{\ell_{2706}}$ dt$^{\ell_{2705}}$;} ;
 $\mathbf{require\_nsb(xJupiter,11)}$$^{\ell_{2710}}$;
 $\mathbf{require\_nsb(xSaturn,11)}$$^{\ell_{2716}}$;
              [...]
  \end{lstlisting}
\hspace{0.2cm}
\tt
  \begin{lstlisting}[mathescape]
 days_per_year$\color{blue}{|56|}$ = 365.24$\color{blue}{|56|}$;
 dt$\color{blue}{|56|}$ = 0.01$\color{blue}{|56|}$;
 t$\color{blue}{|54|}$ = 0.0$\color{blue}{|54|}$;
 t_max$\color{blue}{|53|}$ = 1000.0$\color{blue}{|53|}$;
              [...]
 xJupiter$\color{blue}{|59|}$ = 4.8414316$\color{blue}{|59|}$;
 vxJupiter$\color{blue}{|61|}$ = 0.0016600767$\color{blue}{|61|}$
 *$\color{blue}{|61|}$ days_per_year$\color{blue}{|61|}$;
 massJupiter$\color{blue}{|55|}$ = 9.5479196E-4$\color{blue}{|55|}$
 *$\color{blue}{|55|}$ solar_mass$\color{blue}{|55|}$;
 xSaturn$\color{blue}{|58|}$ = 8.343367$\color{blue}{|58|}$;
              [...]
 vxSaturn$\color{blue}{|61|}$ = -0.002767425$\color{blue}{|61|}$
 *$\color{blue}{|61|}$ days_per_year$\color{blue}{|61|}$;
 massSaturn$\color{blue}{|53|}$ = 2.8588597E-4$\color{blue}{|53|}$
 *$\color{blue}{|53|}$ solar_mass$\color{blue}{|53|}$;
              [...]
 while (t < t_max) {
   dx$\color{blue}{|46|}$ = xJupiter$\color{blue}{|46|}$ -$\color{blue}{|46|}$ xSaturn$\color{blue}{|47|}$;
   dy$\color{blue}{|45|}$ = yJupiter$\color{blue}{|45|}$ -$\color{blue}{|45|}$ ySaturn$\color{blue}{|46|}$;
   dz$\color{blue}{|44|}$ = zJupiter$\color{blue}{|44|}$ -$\color{blue}{|44|}$ zSaturn$\color{blue}{|45|}$;
   distance$\color{blue}{|44|}$ = sqrt(dx$\color{blue}{|46|}$ *$\color{blue}{|46|}$
   dx$\color{blue}{|46|}$ +$\color{blue}{|45|}$ dy$\color{blue}{|45|}$ *$\color{blue}{|45|}$
   dy$\color{blue}{|45|}$ +$\color{blue}{|44|}$ dz$\color{blue}{|35|}$ *$\color{blue}{|35|}$ dz$\color{blue}{|35|}$)$\color{blue}{|44|}$;
   mag$\color{blue}{|44|}$ = dt$\color{blue}{|44|}$ /$\color{blue}{|44|}$ distance$\color{blue}{|44|}$ *$\color{blue}{|44|}$
   distance$\color{blue}{|44|}$ *$\color{blue}{|44|}$ distance$\color{blue}{|44|}$;
   vxJupiter$\color{blue}{|58|}$ = vxJupiter$\color{blue}{|59|}$ -$\color{blue}{|58|}$
   dx$\color{blue}{|41|}$ *$\color{blue}{|41|}$ massSaturn$\color{blue}{|41|}$ *$\color{blue}{|41|}$ mag$\color{blue}{|41|}$;
              [...]
   vxSaturn$\color{blue}{|59|}$ = vxSaturn$\color{blue}{|60|}$ +$\color{blue}{|59|}$
   dx$\color{blue}{|43|}$ *$\color{blue}{|43|}$ massJupiter$\color{blue}{|43|}$ *$\color{blue}{|43|}$ mag$\color{blue}{|43|}$;
              [...]
   xJupiter$\color{blue}{|53|}$ = xJupiter$\color{blue}{|54|}$ +$\color{blue}{|53|}$
   dt$\color{blue}{|47|}$ *$\color{blue}{|47|}$ vxJupiter$\color{blue}{|47|}$;
   xSaturn$\color{blue}{|53|}$ = xSaturn$\color{blue}{|54|}$ +$\color{blue}{|53|}$
   dt$\color{blue}{|46|}$ *$\color{blue}{|46|}$ vxSaturn$\color{blue}{|46|}$;
              [...]
   t$\color{blue}{|53|}$ = t$\color{blue}{|54|}$ +$\color{blue}{|53|}$ dt$\color{blue}{|38|}$;};
 $\mathbf{require\_nsb(xJupiter,11)}$;
 $\mathbf{require\_nsb(xSaturn,11)}$;
             [...]
  \end{lstlisting}
\end{tabular}
\vspace{0.2cm}
\hrule
\vspace{0.2cm}
\caption{\label{running} Left: source program annotated with labels. Right: program with POP generated data types with ILP formulation.}
\end{figure}
 Our tool, POP, automates precision tuning of floating-point programs.  We reduce the problem of precision tuning to determining  which  program  variables,  if  any,  can  have their  types  changed  to  a  lower  precision  while satisfying the user accuracy assertions. In this section, we introduce the gravitational planetary problem simulation code. For this example, we aim at modelling the simulation of a dynamical system describing the orbits of planets in the solar system interacting with each other gravitationally as shown in Figure~\ref{orbites} (note that to for the sake of clarity of the graphic, Figure~\ref{orbites} uses different simulation times for each body.)

We present, in Figure~\ref{running}, excerpts of code that measure the distance between the two planets \texttt{Jupiter} and \texttt{Saturn}. We assume that each body has its own mass (e.g. \texttt{massJupiter}, \texttt{massSaturn}), position (e.g. [\texttt{xJupiter}, \texttt{yJupiter}, \texttt{zJupiter}], [\texttt{xSaturn}, \texttt{ySaturn}, \texttt{zSaturn}]) and velocity (e.g. [\texttt{vxJupiter},\\ \texttt{vyJupiter}, \texttt{zyJupiter}], [\texttt{vxSaturn}, \texttt{vySaturn}, \texttt{vzSaturn}]). Moreover, we suppose that all variables, before POP analysis, are in double precision and that a range determination is performed by dynamic analysis on the program variables (we plan to use a static analyzer in the future). POP assigns to each node of the program's syntactic tree a unique control point in order to determine easily the number of significant bits of the result as mentioned in the left hand side corner of Figure~\ref{running}. Some notations can be stressed about the structure of POP code. We annotate each variable with its unique control point as we can observe in the left hand side program of Figure~\ref{running}, e.g. \texttt{xJupiter$^{\ell_{39}}$ = 4.8414316$^{\ell_{38}}$} denotes that the variable \texttt{xJupiter} has the control point \texttt{$\ell_{39}$} and assigned to the value $4.8414316$ at control point \texttt{$\ell_{38}$}. Considering that $\accd$ denotes the number of significant bits of the variables, the statements \texttt{require\_nsb(xJupiter,11)$^{\ell_{2710}}$} and \texttt{require\_nsb(xSaturn,11)$^{\ell_{2716}}$} on the last two lines of the code inform the system that POP user wants to
have $11$ accurate binary digits on variables \texttt{xJupiter} and \texttt{xSaturn} at their control points \texttt{$\ell_{2710}$} and \texttt{$\ell_{{2716}}$}, respectively.  Note that a result has $n$ significants if the relative error between the exact and approximated results is less than $2^n$.

The key feature of our approach is to generate a set of constraints for each statement of our program (more details in Section~\ref{sec4}). In other words, the accuracy of the arithmetic expressions assigned to variables is determined by semantic equations, in function of the accuracy of the operands. Consider the program of the right hand side of Figure~\ref{running}. We display the POP output N-body program coupled with the generated data types. For a user requirements of $11$ bits on variables \texttt{xJupiter} and \texttt{xSaturn}, POP tunes successfully a large part of the variables of the program (number of constraints solved by GLPK $\simeq$ 3160 with $2468$ variables). For instance, the result of the measured distance between $Jupiter$ and $Saturn$, on line $22$ of the right hand side code of Figure~\ref{running}, is computed with $44$ bits at bit level. Note that the full code
contains other $\accd$ requirements for the other bodies. The $\accd$ given in the right hand side of Figure~\ref{running} are greater than the $\accd$ required on the final results since they have been skewed to ensure
the precision of the whole code at any iteration. Let us also note that even if the computed $\accd$
do not correspond to IEEE74 formats \cite{IEEE754}, one may either take the IEEE754 format immediately above
the computed $\accd$ or choose a multiple precision library such as MPFR \cite{mpfr} or POSIT \cite{UFD19}
(we will discuss more about this point later in this article).

In the next section, we detail the ILP and the PI formulations of the precision tuning problem implemented in POP. Also, we present the nature of the constraints generated for the N-body problem and consequently the new data types already discussed. 

%% file: pop.tex
 \section{Overview of POP}\label{sec4}
\begin{figure}[tb]
\small
\noindent\rule{12.4cm}{0.25mm}
\centering
        \includegraphics[width=10.8cm]{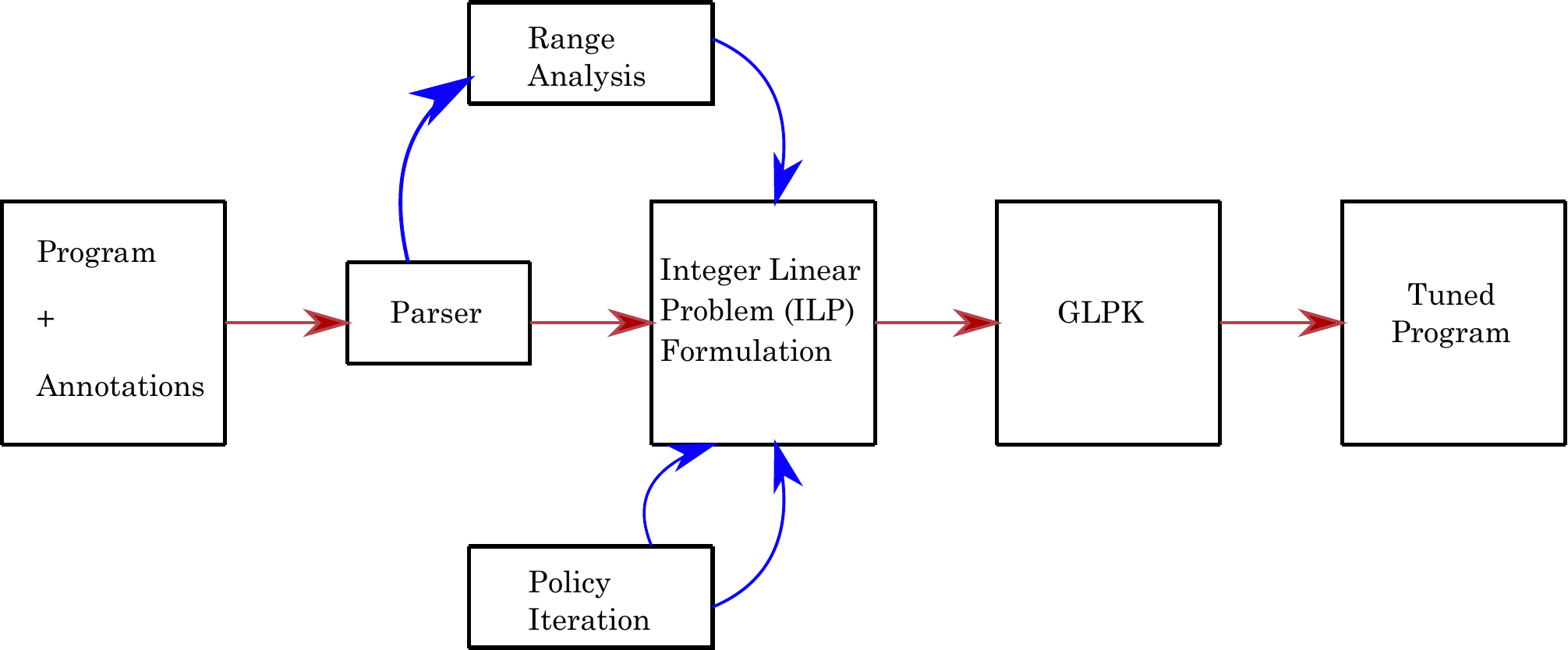}
\noindent\rule{12.4cm}{0.25mm}
\caption{POP overview.}\label{POPoverview}
\end{figure}
POP has been extended in several ways since its first introduction in~\cite{KM19,KM20,KM21,KMA19}. It supports the four elementary operations, trigonometric functions, the square root function, loops, conditionals and arrays. Originally, the analysis in POP were expressed as a set of first order logical propositions among relations between linear integer expressions. Next, these constraints are checked by the Z3 SMT solver~\cite{MB08} in order to return a solution with a certain weight expressing the number of significant bits of the variables.
In the most recent version of POP, the intuition is to use no longer the non-optimizing Z3 SMT solver coupled to a binary search. By that means, we reduce the problem of determining the lowest precision on variables and intermediary values in programs to an Integer Linear Problem (ILP) which can be optimally solved in one breath by a classical linear programming solver as depicted in Figure~\ref{POPoverview}.

Our technique is independent of a particular computer arithmetic. In practice, we handle numbers for which we know their unit in the first place $\ufp$ and their number of significant bits $\accd$ defined as follows. Recall that $\accd$ stands for the number of significant bits of a number. The $\ufp$  of a number $x$ is given in Equation~(\ref{ufp}). This function is used further to describe the way roundoff errors are propagated across computations.
  \begin{equation}\small
     \label{ufp}
       \ufp(x) = \min \{i \in \mathbb{Z} : 2^{i+1} > x \} = \lfloor  \log_2(x) \rfloor \enspace.
     \end{equation}

Noting that the $\ufp$ of the variable values are pre-computed by a prior range analysis. More precisely, the current version of POP performs a dynamic analysis giving an under-approximation of the ranges. Our precision tuning is sensible to the $\ufp$ of the
values. In other words, it is sensible to the order of magnitude of the ranges but not to the exact values. For example, we will obtain the same tuning with the ranges $[3.4,6.1]$ and $[2.5,7.8]$. But, obviously we get a worst tuning if we use a much larger interval, e.g. $[0.0,1000.0]$. In particular, the efficiency of our techniques for loops depends on the precision of the range analysis for loops.

\paragraph{\textbf{Integer Linear Problem Formulation}} In order to explain the obtained data types of our N-body program already illustrated in the right hand corner of Figure~\ref{running}, we present the system of constraints that corresponds to a pure ILP formulation as shown in Equation~(\ref{ilpcstr}). For the sake of conciseness, we will focus on lines $22$ to $24$ that measure the distance between the \texttt{Jupiter} and \texttt{Saturn} bodies (tuned program in the right hand corner of Figure~\ref{running}). To make it easier to follow our reasoning, we rewrite hereafter the statement under discussion annotated with the control points.
\vspace{-0.2em}
\begin{equation*}\begin{array}{rl}\scriptsize\mathtt{distance^{\ell_{788}} = sqrt(}&
         \scriptsize\mathtt{dx^{\ell_{771}} \times^{\ell_{774}} dx^{\ell_{773}}
   +^{\ell_{780}} dy^{\ell_{776}} \times^{\ell_{779}} dy^{\ell_{778}}}\\
   & \scriptsize\mathtt{+^{\ell_{786}} dz^{\ell_{782}}
   \times^{\ell_{785}} dz^{\ell_{784}})^{\ell_{787}}}
   \end{array}\end{equation*}
\normalsize
 \begin{equation}\label{ilpcstr}
\scriptsize
C_1=
\left\{
\begin{array}{l}
\accd(\ell_{780}) \ge \accd(\ell_{786}) + (-7) + \xi(\ell_{786})(\ell_{780}, \ell_{785}) - 7, \\
\accd(\ell_{786}) \ge \accd(\ell_{787}) + (-1) + 0, \\
\accd(\ell_{774}) \ge \accd(\ell_{780}) + 7 + \xi(\ell_{780})(\ell_{774}, \ell_{779})- 7, \\
\accd(\ell_{779}) \ge \accd(\ell_{780})+ 6 + \xi(\ell_{780})(\ell_{774}, \ell_{779})- 7, \\
\accd(\ell_{785}) \ge \accd(\ell_{786})+ (-3) + \xi(\ell_{786})(\ell_{780}, \ell_{785})- 7, \\
\accd(\ell_{787}) \ge\accd(\ell_{788}),
\accd(\ell_{771}) \ge \accd(\ell_{774})+ \xi(\ell_{774})(\ell_{771}, \ell_{773})- 1, \\
                             ...\\
\ \xi(\ell_{780}) (\ell_{774}, \ell_{779})\ge 1, \ \ \xi(\ell_{786})(\ell_{780}, \ell_{785})\ge 1 \\
\ \xi(\ell_{774}) (\ell_{771}, \ell_{773})\ge 1, \ \ \xi(\ell_{779})(\ell_{776}, \ell_{778})\ge 1 \\
\ \xi(\ell_{785}) (\ell_{782}, \ell_{784})\ge 1 \\
\end{array}\right\}
\end{equation}
For this statement, POP generates $17$ constraints as shown in system $C_1$ of Equation~(\ref{ilpcstr}). We assign to each control point (here $771$ to $787$) the integer variable $\accd$  which are determined by solving the system $C_1$. The first two constraints of Equation~(\ref{ilpcstr}) are relative to the $\accd$  of the additions stored at control points $\ell_{780}$ and $\ell_{786}$ respectively. The numbers computed corresponds to the $\ufp$ of the variable values e.g. $\ufp(\ell_{780})= -1$. The following constraints that compute respectively $\accd(\ell_{774})$, $\accd(\ell_{779})$ and $\accd(\ell_{785})$ are generated for the multiplication. The constraint $\accd(\ell_{787}) \ge\accd(\ell_{788})$ is for the square root function. Moreover, the constraint generated for $\accd(\ell_{771})$ is relative to variable $dx$ (same reasoning for variables $dy$ and $dz$ on their control points). Note that POP generates such constraints for all the statements of the N-body program.

Let us now focus on the last five constraints of system $C_1$. We introduce a constant function $\xi$. In the present ILP of Equation~(\ref{ilpcstr}), we assume that $\xi$ is a constant function equal to $1$. This function corresponds to the carry bit that can be propagated at each operation which is expressed by adding a supplementary bit on the elementary operation result. For instance, the constraint $\xi(\ell_{780}) (\ell_{774}, \ell_{779})\ge 1$ indicates that a carry bit is propagated on the result of the addition stored at control point $\ell_{780}$ which is correct but pessimistic (we highlight more the utility of $\xi$ in the next paragraph).
Finally, for a user accuracy requirement of $11$ bits as displayed in Figure~\ref{running}. POP calls the GLPK~\cite{GLPK} solver and consequently finds the least precision needed for all the N-body problem variables as we can observe hereafter ($14636$ is the total number of bits of the whole program after POP optimization).
\begin{equation*}\begin{array}{rl}\scriptsize\mathtt{
distance{|44|} = sqrt(}&\mathtt{dx|46| *|46| dx|46| +|45| dy|45| *|45| dy|45|}\\
&\mathtt{ +|44| dz|35| *|35| dz|35|)}
\end{array}\end{equation*}

\paragraph{\textbf{Policy Iteration to Refine Carry Bit Propagation}}
In the ILP formulation in the above paragraph, we have over-approximated the carry bit function by $\xi =1$. In contrast, this function becomes very
costly in large codes if we perform a lot of computations and therefore the errors would be considerable. For example, if two operands and their errors do not overlap then adding a carry bit is useless. In what follows, we propose an optimization to use a more precise $\xi$ function. Accordingly, when we model this optimization the problem will not remain an ILP any longer, with $\min$ and $\max$ operators that arise, as shown in the refined system of constraints $C_2$ of Equation~(\ref{ipcstr}). Thus, we use the policy iteration method~\cite{CGGMP05} to find an optimal solution.
  \begin{equation}\label{ipcstr}
\scriptsize
C_2=
\left\{
\begin{array}{l}
 \acce(\ell_{780}) \ge  \acce(\ell_{774}), \\
 \acce(\ell_{780}) \ge \acce(\ell_{779}), \\
 \accd(\ell_{780}) \ge 7 - 6 + \accd(\ell_{779})- \accd(\ell_{774})+ \acce(\ell_{779})+\xi(\ell_{780}, \ell_{774}, \ell_{779}), \\
 \acce(\ell_{780}) \ge 6 - 7 + \accd(\ell_{774})- \accd(\ell_{779})+ \acce(\ell_{774})+\xi(\ell_{780}, \ell_{774}, \ell_{779}), \\
 \acce(\ell_{786}) \ge  \acce(\ell_{780}), \\
 \acce(\ell_{786}) \ge \acce(\ell_{785}), \\
 \accd(\ell_{786}) \ge 7 - (-3) + \accd(\ell_{785})- \accd(\ell_{780})+ \acce(\ell_{785})+\xi(\ell_{786}, \ell_{780}, \ell_{785}), \\
 \acce(\ell_{786}) \ge 3 - 7 + \accd(\ell_{780})- \accd(\ell_{785})+ \acce(\ell_{780})+\xi(\ell_{786}, \ell_{780}, \ell_{785}), \\
 \acce(\ell_{774}) \ge  \accd(\ell_{771}) + \acce(\ell_{771})+ \acce(\ell_{773}) - 2, \\
... \\
 \xi(\ell_{780})(\ell_{774}, \ell_{779}) = \min\left(\begin{array}{l}\max\big(6 - 7 + \accd(\ell_{774}) +  \acce(\ell_{774}), 0\big),\\ \max\big(7 - 6 + \accd(\ell_{779}) + \acce(\ell_{779}), 0\big),1 \end{array}\right) \\

 \xi(\ell_{786})(\ell_{780}, \ell_{785}) = \min\left(\begin{array}{l}\max\big(-3 - 7 + \accd(\ell_{780}) +  \acce(\ell_{780}), 0\big),\\ \max\big(7 - (-3) + \accd(\ell_{785}) + \acce(\ell_{785}), 0\big),1 \end{array}\right) \\
\end{array}\right\}
\end{equation}
 Equation~(\ref{ipcstr}) displays the new constraints that we add to the global system of constraints in the case where we optimize the carry bit of the elementary operations. Before introducing these constraints, we define in Equation~(\ref{ulp}) the unit in the last place $\ulp$ of a number $x$.
 \begin{equation}\label{ulp}
 \small
    \ulp(x) = \ufp(x)- \accd(x) + 1 \enspace.
\end{equation}

The principle of the new $\xi$ function is as follows: if the $\ulp$ of one of the two operands (or errors) is greater than the $\ufp$ (see Equation~(\ref{ufp})) of the other one (or conversely) then the two numbers are not aligned and no carry bit can be propagated through the operation (otherwise $\xi$ = 1). Not surprisingly, our new system of constraint $C_2$ introduces a new integer quantity $\acce$ which corresponds to the number of significant bits of the error which needs to be estimated. Formerly, let a number $x$, we define $\ufpe(x)$ and $\ulpe(x)$ as the unit in the first place and in the last place respectively of the error on $x$. From equations~(\ref{ufp}) and (\ref{ulp}), we have  $\ufpe(x) = \ufp(x) - \accd(x)$ and $\ulpe(x) = \ufpe(x) - \acce(x) + 1$ and consequently we can compute $\acce(x)$.

In practice, policy iteration makes it possible to break the $\min$ in the $\xi(\ell_{780})$ $(\ell_{774}, \ell_{779})$ and $\xi(\ell_{786})(\ell_{780}, \ell_{785})$ functions of the two additions as shown in Equation~(\ref{ipcstr}) by choosing the max between the terms. Next, it becomes possible to solve the corresponding ILP. If no fixed point is reached, POP iterates until a solution is found.
By applying this optimization, the new data types of the statement of lines $22$ to $24$ in Figure~\ref{running} are given as follows.
\begin{equation*}\begin{array}{rl}\scriptsize\mathtt{
   distance|41| = sqrt(}&\mathtt{dx|42| *|42| dx|42| +|42| dy|42| *|42| dy|42|}\\
   &\mathtt{ +|41| dz|31| *|31| dz|31|)}
   \end{array}
\end{equation*}

By comparing with the formats already presented with the ILP method, it is obvious the gain of precision that we obtain on each variable and operation of this statement. With the PI method, the total number of bits of the optimized N-body program is $\simeq$ 14335 (a gain of more than $300$ bits compared to the ILP formulation). In term of complexity, for both ILP and PI methods, POP generates a linear number of constraints and variables in the size of the analyzed program and finds the best tuning of the variables in  polynomial time. 

%% file: experim.tex
 \section{Experimental Results}\label{sec5}

 \begin{table}[tb]\scriptsize
\centerline{
\begin{tabular}{ccccccc}
   \texttt{nsb}      &  \texttt{11} & \texttt{18} & \texttt{24} & \texttt{34} & \texttt{43} & \texttt{53}    \\
   \\
\hline
\\
\multicolumn{7}{c}{\textbf{Simulation time: 10 years}}\\
\\
\hline
\\
Jupiter  & $5.542\cdot 10^{-4}$ & $1.650\cdot 10^{-6}$ & $1.577\cdot 10^{-7}$ & $4.998\cdot 10^{-10}$ & $5.077\cdot 10^{-10}$ & $5.076\cdot 10^{-10}$\\
\\
Saturn   & $1.571\cdot 10^{-3}$ & $2.111\cdot 10^{-5}$ & $1.326\cdot 10^{-7}$ & $4.427\cdot 10^{-10}$ & $3.119\cdot 10^{-10}$ & $3.117\cdot 10^{-10}$\\
\\
Uranus   & $2.952\cdot 10^{-3}$ & $2.364\cdot 10^{-5}$ & $1.140\cdot 10^{-7}$ & $3.072\cdot 10^{-10}$ & $7.212\cdot 10^{-11}$ & $7.236\cdot 10^{-11}$\\
\\
Neptune  & $2.360\cdot 10^{-3}$ & $3.807\cdot 10^{-5}$ & $2.206\cdot 10^{-7}$ & $5.578\cdot 10^{-10}$ & $1.751\cdot 10^{-10}$ & $1.757\cdot 10^{-10}$\\
\\
\\
Runtime     & 2'59 & 2'52 & 2'57 & 2'56 & 3'10 & 2'59\\
\\
POP Time    &   25''   & 22''  & 22''  & 24'' & 23'' & 24''  \\
\\
\hline
\\
\multicolumn{7}{c}{\textbf{Simulation time: 30 years}}\\
\\
\hline
\\
Jupiter  & $7.851\cdot 10^{-4}$ & $1.282\cdot 10^{-5}$ & $3.194\cdot 10^{-8}$ & $1.066\cdot 10^{-8}$ & $1.064\cdot 10^{-8}$ & $1.064\cdot 10^{-8}$\\
\\
Saturn   & $3.009\cdot 10^{-3}$ & $1.934\cdot 10^{-5}$ & $2.694\cdot 10^{-7}$ & $1.7477\cdot 10^{-8}$ & $1.777\cdot 10^{-8}$ & $1.777\cdot 10^{-8}$\\
\\
Uranus   & $6.839\cdot 10^{-4}$ & $6.132\cdot 10^{-5}$ & $8.901\cdot 10^{-7}$ & $5.105\cdot 10^{-10}$ & $1.464\cdot 10^{-10}$ & $1.457\cdot 10^{-10}$\\
\\
Neptune  & $2.971\cdot 10^{-3}$ & $2.0227\cdot 10^{-5}$ & $2.469\cdot 10^{-7}$ & $3.869\cdot 10^{-10}$ & $4.775\cdot 10^{-10}$ & $4.779\cdot 10^{-10}$\\
\\
\\
Runtime     & 2'39 & 2'45 & 2'43 & 2'56 & 2'48 & 2'40\\
\\
POP Time    &   38''   & 39''  & 41''  & 37'' & 37''& 37'' \\
\\
\hline
\end{tabular}}\normalsize
\vspace{0.4cm}
\caption{Distances between the exact position (computed with 500 bits) and the position computed
with $n$ bits. Distances given for each body after $10$ and $30$ years of simulation. Followed by POP analysis time and the execution time of the MPFR generated code.}\label{simulation}
\end{table}

\begin{figure}[t]
\begin{center}
\hrule
\tt\scriptsize
\begin{lstlisting}[mathescape]
 xJupiter = $\mathbf{mpfr}$(4.841431617736816,59)
 yJupiter = $\mathbf{mpfr}$(-1.1603200435638428,60)
 zJupiter = $\mathbf{mpfr}$(-0.10362204164266586,57)
 vxJupiter = $\mathbf{mpfr}$($\mathbf{mpfr}$(0.001660076668485999,61)*$\mathbf{mpfr}$(days_per_year,61),61)
 vyJupiter = $\mathbf{mpfr}$($\mathbf{mpfr}$(0.007699011359363794,61)*$\mathbf{mpfr}$(days_per_year,61),61)
 vzJupiter = $\mathbf{mpfr}$($\mathbf{mpfr}$(-6.904600013513118E-5,61)*$\mathbf{mpfr}$(days_per_year,61),61)
 massJupiter = $\mathbf{mpfr}$($\mathbf{mpfr}$(9.547919617034495E-4,55)*$\mathbf{mpfr}$(solar_mass,55),55)
[...]
 $\mathbf{while}$( t<t_max):
    dx = $\mathbf{mpfr}$($\mathbf{mpfr}$(xSun,57)-$\mathbf{mpfr}$(xJupiter,59),58)
    dy = $\mathbf{mpfr}$($\mathbf{mpfr}$(ySun,60)-$\mathbf{mpfr}$(yJupiter,60),57)
    dz = $\mathbf{mpfr}$($\mathbf{mpfr}$(zSun,60)-$\mathbf{mpfr}$(zJupiter,57),55)
    distance = gmpy2.sqrt($\mathbf{mpfr}$($\mathbf{mpfr}$($\mathbf{mpfr}$($\mathbf{mpfr}$(dx,58)*$\mathbf{mpfr}$(dx,58),58)
 +$\mathbf{mpfr}$($\mathbf{mpfr}$(dy,57)*$\mathbf{mpfr}$(dy,57),57),57)+$\mathbf{mpfr}$($\mathbf{mpfr}$(dz,46)
               *$\mathbf{mpfr}$(dz,46),46),56))
    mag = $\mathbf{mpfr}$($\mathbf{mpfr}$(dt,56)/$\mathbf{mpfr}$($\mathbf{mpfr}$($\mathbf{mpfr}$(distance,56)
 *$\mathbf{mpfr}$(distance,56),56)*$\mathbf{mpfr}$(distance,56),56),56)
    vxJupiter = $\mathbf{mpfr}$($\mathbf{mpfr}$(vxJupiter,61)+$\mathbf{mpfr}$($\mathbf{mpfr}$($\mathbf{mpfr}$(dx,56)
 *$\mathbf{mpfr}$(massSun,56),56)*$\mathbf{mpfr}$(mag,56),56),60)
    vyJupiter = $\mathbf{mpfr}$($\mathbf{mpfr}$(vyJupiter,61)+$\mathbf{mpfr}$($\mathbf{mpfr}$($\mathbf{mpfr}$(dy,54)
 *$\mathbf{mpfr}$(massSun,54),54)*$\mathbf{mpfr}$(mag,54),54),60)
    vzJupiter = $\mathbf{mpfr}$($\mathbf{mpfr}$(vzJupiter,61)+$\mathbf{mpfr}$($\mathbf{mpfr}$($\mathbf{mpfr}$(dz,55)
 *$\mathbf{mpfr}$(massSun,55),55)*$\mathbf{mpfr}$(mag,55),55),60)
    [...]
    xJupiter = $\mathbf{mpfr}$($\mathbf{mpfr}$(xJupiter,54)+$\mathbf{mpfr}$($\mathbf{mpfr}$(dt,47)
               *$\mathbf{mpfr}$(vxJupiter,47),47),53)
    yJupiter = $\mathbf{mpfr}$($\mathbf{mpfr}$(yJupiter,54)+$\mathbf{mpfr}$($\mathbf{mpfr}$(dt,47)
               *$\mathbf{mpfr}$(vyJupiter,47),47),53)
    zJupiter = $\mathbf{mpfr}$($\mathbf{mpfr}$(zJupiter,54)+$\mathbf{mpfr}$($\mathbf{mpfr}$(dt,47)
               *$\mathbf{mpfr}$(vzJupiter,47),47),53)
    [...]
}

  \end{lstlisting}
\hrule
\end{center}
\caption{Python MPFR code automatically generated by POP for the N-body problem for a $\accd$ requirement of $18$ bits on the positions of the planets at the end of the simulation.}\label{mpfr}
\end{figure}
     \begin{figure}[tb]
     \begin{center}
          \noindent\rule{12.4cm}{0.25mm}
\begin{subfigure}{0.51\textwidth}
\includegraphics[width=\linewidth]{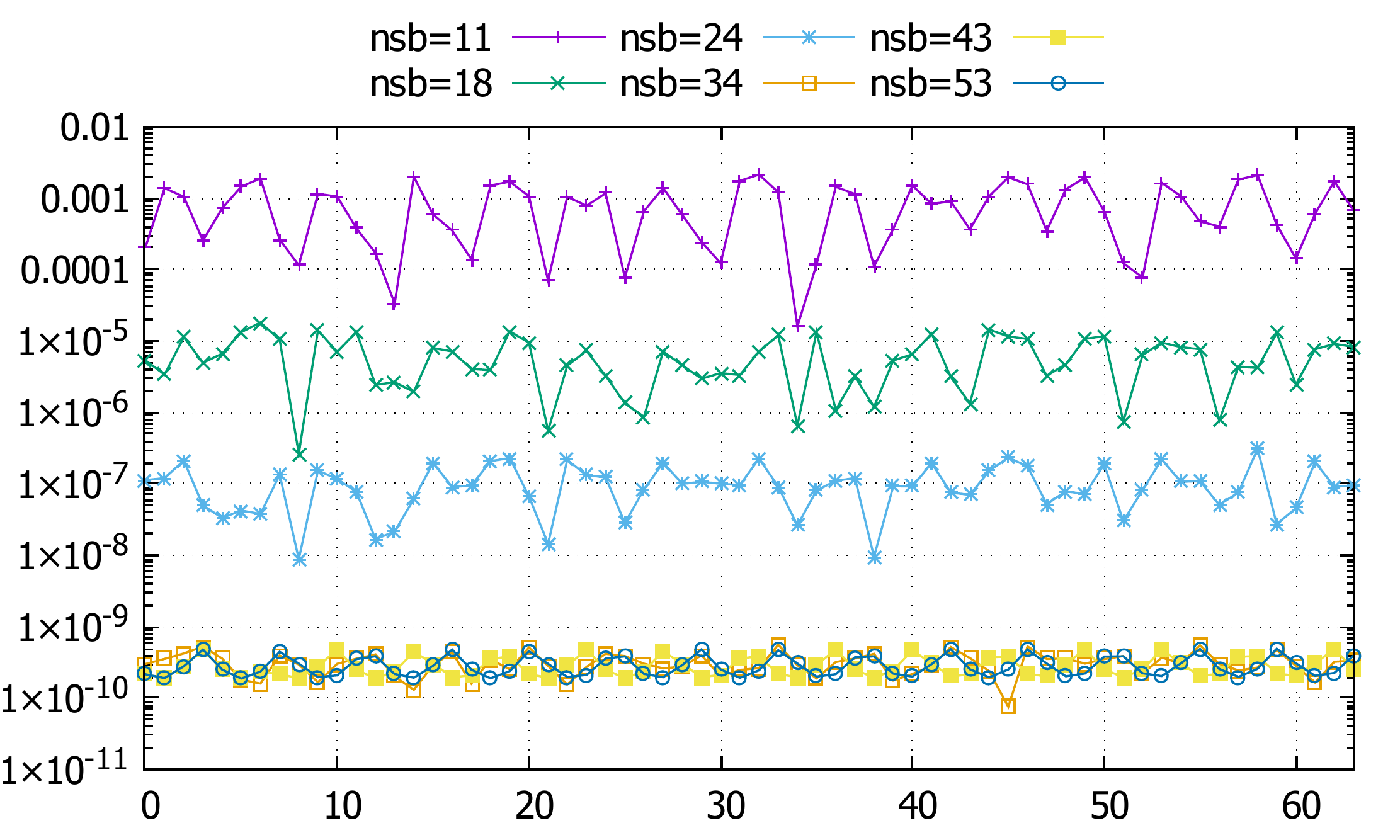}
\caption{Jupiter} \label{distance:jupiter}
\end{subfigure}\hspace*{\fill}
\begin{subfigure}{0.51\textwidth}
\includegraphics[width=\linewidth]{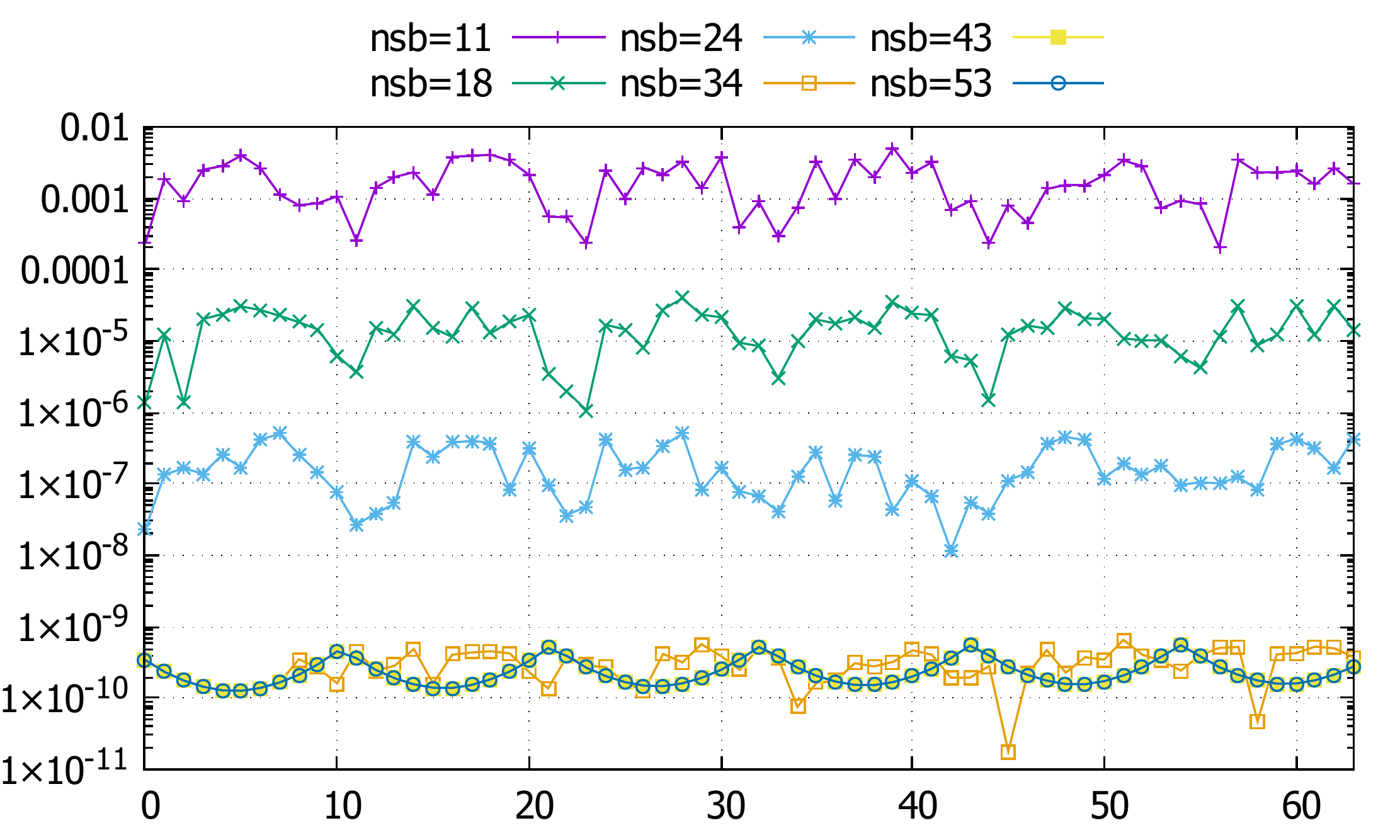}
\caption{Saturn} \label{distance:saturn}
\end{subfigure}
\medskip
\begin{subfigure}{0.51\textwidth}
\includegraphics[width=\linewidth]{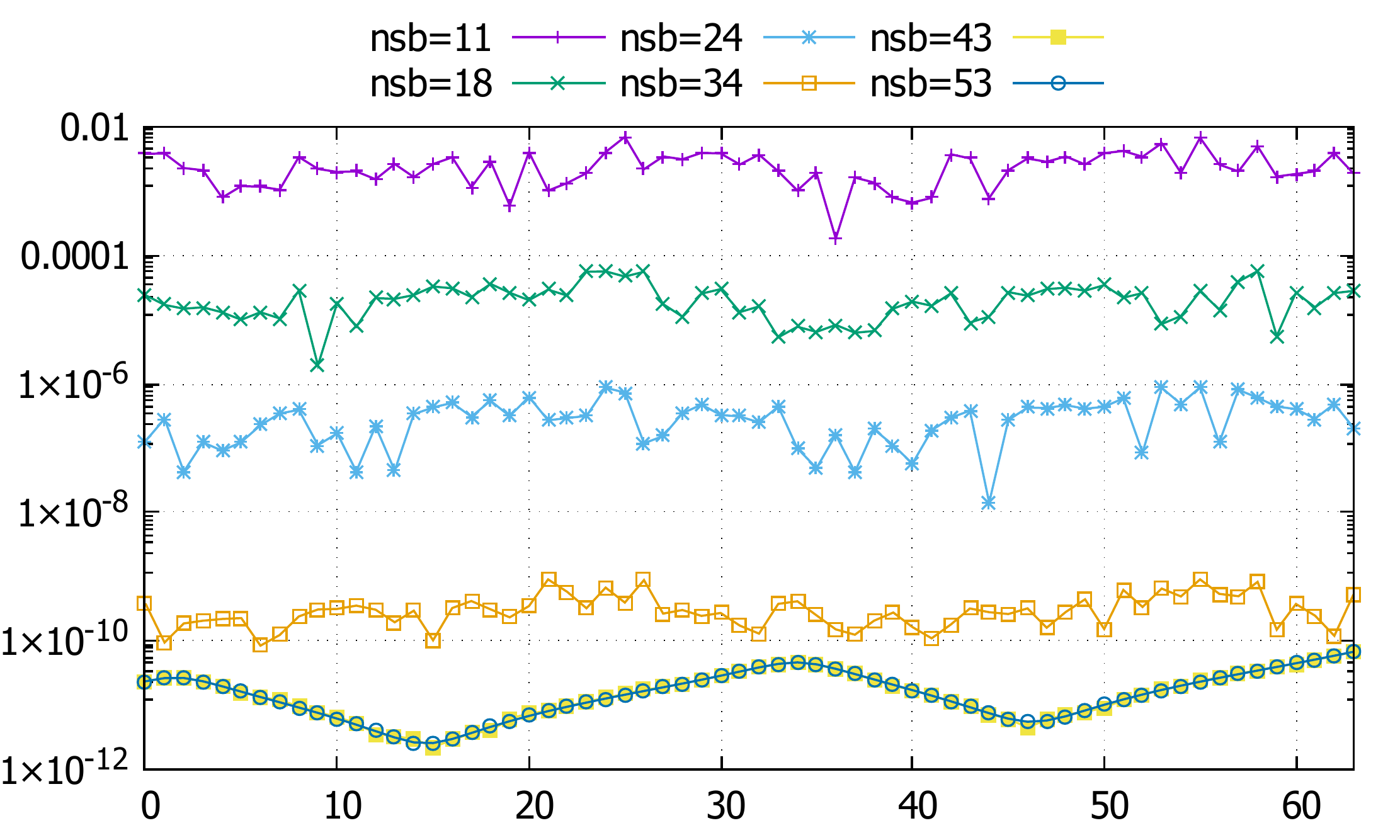}
\caption{Uranus} \label{distance:uranus}
\end{subfigure}\hspace*{\fill}
\begin{subfigure}{0.51\textwidth}
\includegraphics[width=\linewidth]{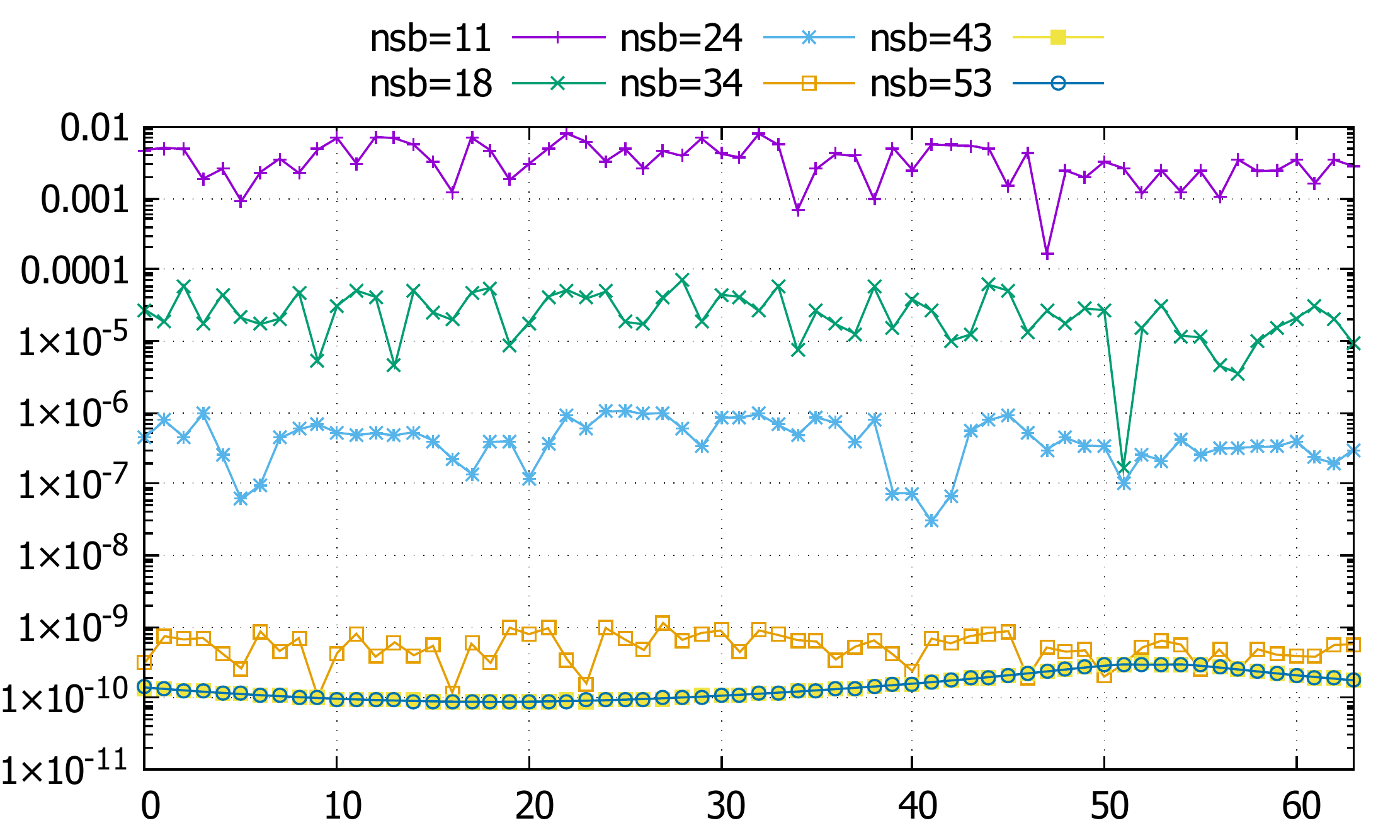}
\caption{Neptune} \label{distance:neptune}
\end{subfigure}
     \noindent\rule{12.4cm}{0.25mm}
\caption{Distance between the exact and the computed position for the $5$ bodies with $11$, $18$, $24$, $34$, $43$ and $53$ bits.} \label{distance}
\end{center}
\end{figure}
In this section, our goal is to evaluate the performances of POP in tuning the code simulating the behaviour of the different bodies of our example. We note that the N-body program has been excerpted (not fully) from~\cite{demeure} which relies on a second order differential equation solved by Euler's method.
 Now, we shed some light on the POP tool outline already depicted in Figure~\ref{POPoverview}. POP has been developed in JAVA. It uses the ANLTR v$4$.$7$.$1$\footnote{\texttt{https://www.antlr.org/}} framework to parse the different input programs. As mentioned in Section~\ref{sec4}, we reduce the precision tuning problem to an ILP by generating a set of semantical equations which can be solved by a linear solver. The integer solution to this problem, computed in polynomial time by a (real) linear programming solver, we use GLPK v$4$.$65$~\cite{GLPK}, gives the optimal data types at the bit level.

We ran our precision tuning analysis on the N-body problem with different $\accd$ requirements on the program variables: $11$, $18$, $24$, $34$, $43$ and $53$ bits.
This shows the ability of POP to tune programs in function of the IEEE754 formats ($11$, $24$, $53$) \cite{IEEE754} as
well as for arbitrary word length which can be encoded using libraries such as MPFR \cite{mpfr} or POSIT~\cite{UFD19}.
We test the efficiency of POP analysis in several ways. The experiments shown in Table~\ref{simulation} seek to measure the distances between the exact position of each of the bodies of our planetary system and the position computed with an $\accd$ of $11$, $18$, $24$, $34$, $43$ and $53$ bits.
The distances presented in Table~\ref{simulation} are given for a single position on the planets which follow the orbits previously presented in Figure~\ref{orbites}. The positions are taken after  of $10$ and $30$ years of simulation time.

More precisely, for this experimentation, we generate the N-body program with all computations done on $500$ bits
(we assume that this gives the exact solution)
and we also generate by the same manner an MPFR~\cite{mpfr} code with the optimized data types returned by POP. For example, as we can observe in Table~\ref{simulation}, for an $\accd = 11$, the distance measured for \texttt{Jupiter} is of the order of $10^{-4}$ for $10$ years of simulation which confirms the usefulness of our analysis: desirable results (also for the remaining planets) that respects the user $\accd$ requirement where the worst error is of $2^{-11}$ for $\accd$ = $11$. For a simulation of $10$ and $30$ years, the runtime spent to measure these distances reaches maximally $2$ minutes $59$ seconds for an $\accd$ = $53$. Concerning the POP time, our analysis took as little as $25$ seconds ($\accd=11$) to find that we can lower the precision of the majority of variables of the N-body program for a simulation time of $10$ years and does not exceed $41$ seconds for a simulation time of $30$ years ($\accd$ = $24$). With this speed, we believe that for large codes POP achieves its best tuning in a minimal time. Figure~\ref{mpfr} depicts the capability of POP to generate automatically a Python MPFR version of the N-body program on the position of the planets at the end of the simulation. The MPFR code is annotated with the optimized formats returned by POP after analysis for $\accd = 18$. In the future, we plan to also generate code for
libraries based on the POSIT number system\footnote{https://github.com/stillwater-sc/universal}.

We end this section by focusing on the curves of Figure~\ref{distance}. For this experiment, we plot the distance between the exact and the computed position for each body at each instant of the simulation. This extends the results of Table~\ref{simulation} to all instants and not to specific ones. Consequently, we deduce from these observations that the measured error is controlled for the different planets at each iteration of the simulation. 

%% file: conclu.tex
 \section{Concluding Remarks}\label{sec6} \vspace{-0.2em}

The primary goal of our work was to provide a new approach for mixed-precision tuning, totally different from the existing ones. The novelty of our technique is to propose a semantical modelling of the propagation of the numerical errors throughout the code expressed as a set of constraints. We have defined two variants of methods. The first one corresponds to a pure ILP with an over-approximation of the carries in the elementary operations.
The second one aims to use a more precise carry bit function and is solved by the policy iteration technique~\cite{CGGMP05}. Both two methods have been implemented in our tool POP. We believe this static analysis performed by our automated tool is unique. The effectiveness of POP has already been demonstrated on a variety of programs coming from different fields.

In this article, we have shown that POP is able to tune the N-body program according to different number of significant bit required by the user. The results presented are promising in term of the analysis technique, speed and efficiency. The only limitation we can face is the size of the problem accepted by the solver.
In addition, we have also shown that POP is able to generate code for multiple precision libraries, MPFR in practice, and we plan
to integrate POSIT libraries in the near future.

Broadly speaking, our important future directions include handling Deep Neural Network's (DNNs) for which saving resources is essential. Also, code synthesis
for the fixed-point arithmetic and assigning the same precision to pieces of code are perspectives we aim at explore at short term. 